# Experimental Demonstration of Compensation of Beam-Beam Effects by Electron Lenses

*Vladimir Shiltsev, Yuri Alexahin, Vsevolod Kamerdzhiev,*

*Gennady Kuznetsov, and Xiao-Long Zhang*

*Fermi National Accelerator Laboratory, PO Box 500, Batavia, IL 60510, USA*

*Kip Bishofberger*

*Los Alamos National Laboratory, Los Alamos, NM 87545, USA*

*Abstract*

We report the first experimental demonstration of compensation of beam-beam interaction effects in high-energy particle collider by using space-charge forces of a low-energy electron beam. In our experiments, an electron lens, a novel instrument developed for the beam-beam compensation, was set on a 980-GeV proton bunch in the Tevatron proton-antiproton collider. The proton bunch losses due to its interaction with antiproton beam were reduced by a factor of 2 when the electron lens was operating. We describe the principle of electron lens operation and present experimental results.

The luminosity of storage ring colliders is limited by the effects of electromagnetic (EM) interaction of one beam on the particles of the other beam which leads to a blowup of beam sizes, a loss of beam intensities and unacceptable background rates in the high energy physics (HEP) detectors. This beam-beam interaction is parameterized by a dimensionless *beam-beam parameter* $\xi \equiv r_0 N / 4\pi\varepsilon$, where $r_0 = e^2/mc^2$ denotes the particle's classical radius, $N$ is the number of particles in the opposing bunch and $\varepsilon$ is its the rms normalized emittance related to transverse rms beam size $\sigma$ at the interaction point (IP) as $\varepsilon = \gamma\sigma^2/\beta^*$, $\gamma \gg 1$ is relativistic gamma factor, $\beta^*$ is the beta-function at the IP (for simplicity here, we consider a collider with round Gaussian beams). This dimensionless parameter is equal to the shift of the betatron oscillation tune $Q = f_\beta/f_0$ of core particles due to beam-beam forces. While core particles undergo a significant tune shift, halo particles with large oscillation amplitudes experience negligible tune shift. The EM forces drive nonlinear resonances which can result in instability of particle motion and loss. The beam-beam limit in modern hadron colliders is at $\xi^{max} \cdot N_{IP} \approx 0.01 - 0.02$ ($N_{IP}$ is the number of IPs), while it can exceed $\xi^{max} \cdot N_{IP} \approx 0.1$ in high energy electron-positron colliders [1].

Operation with a greater number of bunches allows a proportional increase of luminosity but requires careful spatial separation of two beams everywhere except at the main IPs. *Long-range* (as opposed to *head-on*) EM interactions of separated beams are also nonlinear and contribute to the limit on collider performance. These long-range effects usually vary from bunch to bunch, making their treatment even more cumbersome.

Besides the technique of electron lenses, the subject of this Letter, there are few beam-beam compensation (BBC) schemes tested experimentally. The 0.8-GeV DCI storage ring at the Laboratoire de l'Accelerateur Lineaire (Orsay, France) had four colliding beams – one positron and one electron coming from each direction. Full space charge and current compensation could be achieved if all the beams had the same intensities and dimensions, but the observed beam-beam limit was not significantly different than with just two beams [2]. These results are attributed to strong coherent beam-beam effects which are characterized by rapid correlated variations of the beam distributions – see Ref.[3] and references therein. Octupole magnets were used for compensation of the cubic nonlinearity in the beam-beam force at the VEPP-4 *e+e-* collider (Novosibirsk, Russia) [4]. Although a several-fold reduction of electron halo loss rate was demonstrated at optimal octupole current, the technique has not found wide application because its efficiency is strongly dependent on the machine tune. Compensation of the EM fields of *separated* beams by placing a current conducting wire at the same distance to the beam as opposite beam was proposed in [5]. Some 20% reduction of the *e+* loss rate due to such a method was observed at the DAFNE (Frascatti, Italy) [6]. The wire-compensation technique is less efficient if multiple beam-beam interactions occur at different distances and betatron phases, and, of course, it is useless for head-on BBC.

Electron lenses were proposed for compensation of both long-range and head-on beam-beam effects in the Fermilab's Tevatron collider (Batavia, USA) [7]. The lens employs a low energy $\beta_e = v/c \ll 1$ beam of electrons which collides with the high-energy bunches over an extended length $L_e$. Electron space charge forces are linear at distances smaller than the characteristic beam radius $r < a_e$ but scale as $1/r$ for $r > a_e$. Correspondingly, such a lens can be used for linear and nonlinear force compensation depending on the beam-size ratio $a_e/\sigma$ and the current-density distribution $j_e(r)$. Main advantages of the electron lens compensation are: a) the electron beam acts on high-energy beams only through EM forces (no nuclear interaction), eliminating radiation issues; b) fresh electrons interact with the high-energy particles each turn, leaving no possibility for coherent instabilities; c) the electron current profile (and thus the EM field profiles) can easily be changed for different applications; d) the electron-beam current can be adjusted between each of the bunches, equalizing the bunch-to-bunch differences and optimizing the performance of all of the bunches in multi-bunch colliders.

Two Tevatron Electron Lenses (TELs) were built and installed in two different locations of the Tevatron ring, A11 and F48. Fig.1 depicts a general layout of the

TELs. The TEL and relevant Tevatron parameters are given in Table I. In order to keep electron beam straight and its distribution unaffected by its own space-charge and main beam EM fields, the electron beam is immersed in a strong magnetic field - about 3 kG at the electron-gun cathode and some 30 kG inside the main superconducting (SC) solenoid. The deviations of the magnetic field lines from a straight line are less than ±100 µm over the entire length of the SC solenoid. The electron beam, following the field lines, therefore does not deviate from the straight Tevatron beam trajectory by more than 20% of the Tevatron beam rms size $\sigma \approx 0.5 - 0.7$ mm in the location of the TELs.

The electron beam's transverse alignment on the proton or antiproton bunches (within 0.2–0.5 mm all along the interaction length) is crucial for successful BBC. The electron beam steering is done by adjusting currents in superconducting dipole correctors installed inside the main solenoid cryostat. It was also important that electron gun generates electron current distribution with wide flat top and smooth radial edges. Such a distribution is generated in the 7.5-mm radius convex cathode electron gun with an optimized electrode geometry [8]. The TEL magnetic system compresses the electron-beam cross-section area in the interaction region by the factor of $B_{main}/B_{gun} \approx 10$ (variable from 2 to 30), proportionally increasing the current density of the electron beam in the interaction region. The electron beam radius $a_e$ in the interaction section (main solenoid) is about 3-4 times the rms size of the

Tevatron beam. Most current experiments have not required more than 0.6 A, though previous tests up to 3.0 A have been performed. In order to enable operation on a single bunch in the Tevatron with bunch spacing of 396 ns, the anode voltage, and consequently the beam current, are modulated with a characteristic on-off time of about 0.6 μs and a repetition rate equal to the Tevatron revolution frequency of $f_0 = 47.7$ kHz by using a HV Marx pulse generator or a HV RF tube base amplifier. The electron pulse timing jitter is less than 1 ns and the peak current is stable to better than one percent, so, the TEL operation does not incur any significant emittance growth More detailed description of the TEL magnetic and electron beam systems, beam diagnostics and tests can be found in Ref. [9] and references therein.

The high-energy protons are focused by the TEL and experience a positive betatron tune shift given by [7]:

$$dQ_{x,y} = +\frac{\beta_{x,y} L_e r_p}{2\gamma ec} \cdot j_e \cdot \left(\frac{1-\beta_e}{\beta_e}\right) \qquad (1).$$

Fig.2 presents results of the measurements of the vertical tune shift $dQ_y$ of 980-GeV protons versus electron current in the TEL installed at the A11 location with a vertical beta-function of $\beta_y = 150\text{m}$, in good agreement with Eq.(1) using the values of $\beta_e = 0.14$ for 5-keV electron energy and $j_e \equiv J_e/\pi a_e^2 = 0.05 \text{A}/\text{mm}^2$ for a 0.6-A beam with an effective radius of about 2 millimeters.

One of the most detrimental effects of the beam-beam interaction in the Tevatron is the significant attrition rate of protons due to their interaction with the antiproton bunches, both in the main IPs and in the numerous long-range interaction regions [10]. The effect is especially large at the beginning of the HEP stores where the total proton beam-beam tune shift induced by antiprotons at two IPs (B0 and D0) can reach the values of $2\xi^{proton} = 0.018$. Fig. 3 shows a typical distribution of proton loss rates at the beginning of an HEP store. Bunches #12, 24, and 36 at the end of each bunch train typically lose about 9% of their intensity per hour while other bunches lose only 4% to 6% per hour. These losses are a very significant part of the total luminosity decay rate of about 20% per hour (again, at the beginning of the high luminosity HEP stores). The losses due to inelastic proton-antiproton interactions at the two main IPs are small (1–1.5%/hr) compared to the total losses. In the Tevatron, 36 bunches in each beam are arranged in three trains of 12 bunches separated by 2.6 µs long abort gaps. Fig.3 shows a large bunch-to-bunch variations in the beam-beam induced proton loss rates within each bunch train but very similar rates for equivalent bunches, e.g. bunches #12, 24, and 36.

In the BBC demonstration experiment, we centered and timed the electron beam of the A11 TEL onto bunch #12 without affecting any other bunches. When the TEL peak current was increased to 0.6A, the lifetime $\tau = N/(dN/dt)$ of bunch #12 went up to 26.6 hours from about 12 hours - see Fig.4. At the same time, the lifetime of bunch #36, an equivalent bunch in the third bunch train, remained low and did not change significantly (at 13.4 hours lifetime). When the TEL current was turned off for fifteen minutes, the lifetimes of both bunches were, as expected, nearly identical (16 hours). The TEL was then turned on again, and once again the lifetime for bunch #12 improved significantly to 43 hours while bunch #36 stayed poor at 23.5 hours. This experiment demonstrates a factor of two improvement in the proton lifetime due to compensation of beam-beam effects with the TEL.

The proton lifetime, dominated by beam-beam effects, gradually improves and reaches roughly 100 hours after 6-8 hours of collisions; this is explained by a decrease in antiproton population and an increase in antiproton emittance, both contributing to a reduction of the beam-beam parameter $\xi^{\text{proton}}$. To study the effectiveness of BBC later in the store, the TEL was repeatedly turned on and off every half hour for 16 hours, again on bunch #12. The relative improvement, defined as the ratio of the proton lifetime with the TEL and without, is plotted in Fig.5. The first two data points correspond to $J_e = 0.6$ A (as is Fig.4 and the above description), but subsequent points were taken with a reduced peak electron current of $J_e = 0.3$ A. This reduction led to a drop of the relative improvement from a factor of 2.03 to a factor of 1.4. A gradual decrease in the relative improvement is visible until after about ten hours, where the ratio reaches 1.0 (no gain in lifetime). At this point, the beam-beam effects have become very small, providing little to compensate. Similar experiments in several other stores with initial luminosities ranging from $1.5 \cdot 10^{32}$ cm$^{-2}$ s$^{-1}$ to $2.5 \cdot 10^{32}$ cm$^{-2}$ s$^{-1}$ repeated these results.

The lifetime improvement due to the TEL can be explained by the positive shift of vertical tune of protons $dQ_y \approx 0.0015$ which makes the detrimental effects of the 12$^{th}$ order resonance $Q_y=7/12=0.583$ weaker. The average Tevatron proton tune $Q_y=0.589$ (which is carefully optimized to minimize overall losses) is just above this resonance, and the bunches at the end of each train, which have lower vertical tunes $Q_y=0.587$ due to a unique beam-beam interaction with antiproton bunches, are subject to stronger beam-beam effects [10]. The TEL moves those protons away from the resonance, thus, resulting in significant reduction of the losses.

In conclusion, we have developed an electron lens - a new instrument for compensation of beam-beam effects in high-energy colliders. Such a lens significantly and reliably improves the lifetime of the Tevatron proton bunches. The observed improvement of the proton lifetime at the beginning of HEP store, when beam brightness and luminosity are the highest and the beam-beam interaction is the strongest, has been as large as a factor of two. Ten hours into a store, the beam-beam effects, and therefore the utility of BBC, decrease significantly. Currently, the TELs are being incorporated into the daily operation of the Tevatron collider.

The versatility of electron lenses allows their use in many other applications, also. For example, the TEL installed at F48 location in the Tevatron for several years is used for removing unwanted DC beam particles out of the Tevatron abort gaps between the bunch trains [11]. Several other electron lens concepts have been proposed for space-charge compensation in high intensity proton synchrotrons [12], reduction of a tune spread in colliding beams [13], and beam collimation in the LHC [14].

We would like to thank A.Burov, V.Danilov, B.Drendel, D.Finley, R.Hively, A.Klebaner, S.Kozub, M.Kufer, L.Tkachenko, J.Marriner, V.Parkhomchuk, H.Pfeffer, V.Reva, A.Seryi, D.Shatilov, N.Solyak, M.Tiunov, A.Valishev, D.Wildman, D.Wolff, and F.Zimmermann for their help, technical contributions, assistance during beam studies, and fruitful discussions on this subject. Fermilab is operated by Fermi Research Alliance Ltd. under Contract No. DE-AC02-76CH03000 with the United States Department of Energy.

TABLE I. Electron Lens and Tevatron Collider parameter list.

| Parameter | Symbol | Value | Unit |
|---|---|---|---|
| *Tevatron Electron Lens* | | | |
| Electron beam energy (oper./max) | $U_e$, | 5/10 | kV |
| Peak electron current (oper./max) | $J_e$ | 0.6/2.3 | A |
| Magnetic field in main/gun solenoid | $B_{main}/B_{gun}$ | 30.1/2.9 | kG |
| $e$-beam radius in main solenoid | $a_e$ | 2.3 | mm |
| Cathode radius | $a_c$ | 7.5 | mm |
| $e$-pulse width, "0-to-0" | $T_e$ | ~600 | ns |
| $e$-pulse repetition rate | $f_0 = c/C$ | 47.7 | kHz |
| Effective interaction length | $L_e$ | 2.0 | m |

*Tevatron Collider Parameters*

| | | | |
|---|---|---|---|
| Circumference | $C$ | 6.28 | km |
| Proton/antiproton beam energy | $E$ | 980 | GeV |
| Proton/antiproton bunch intensity | $N_p / N_a$ | ~250/50-90 | $10^9$ |
| Emittance proton/antiproton (norm., rms) | $\varepsilon_p / \varepsilon_a$ | 17/8 | μm |
| Number of bunches | $N_B$ | 36 | |
| Initial luminosity | $L_0$ | 1.5-2.9 | $10^{32}$ cm$^{-2}$s$^{-1}$ |
| Bunch spacing | $T_b$ | 396 | ns |
| Beta functions at A11 (F48) TEL | $\beta_y / \beta_x$ | 150/68(29/104) | m |

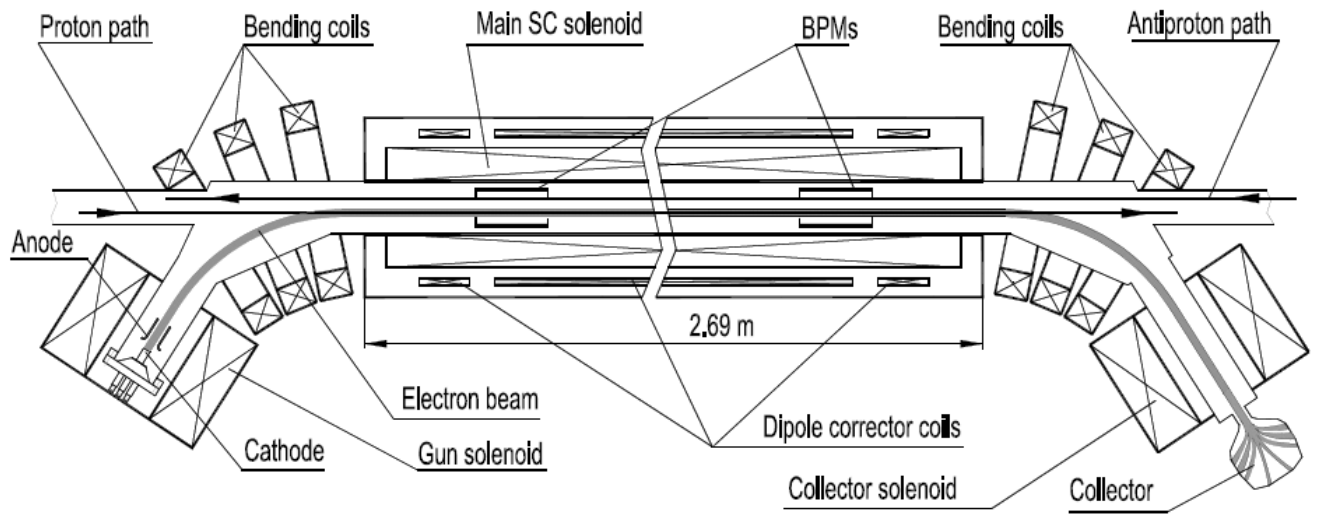

FIG. 1. Layout of the Tevatron Electron Lens.

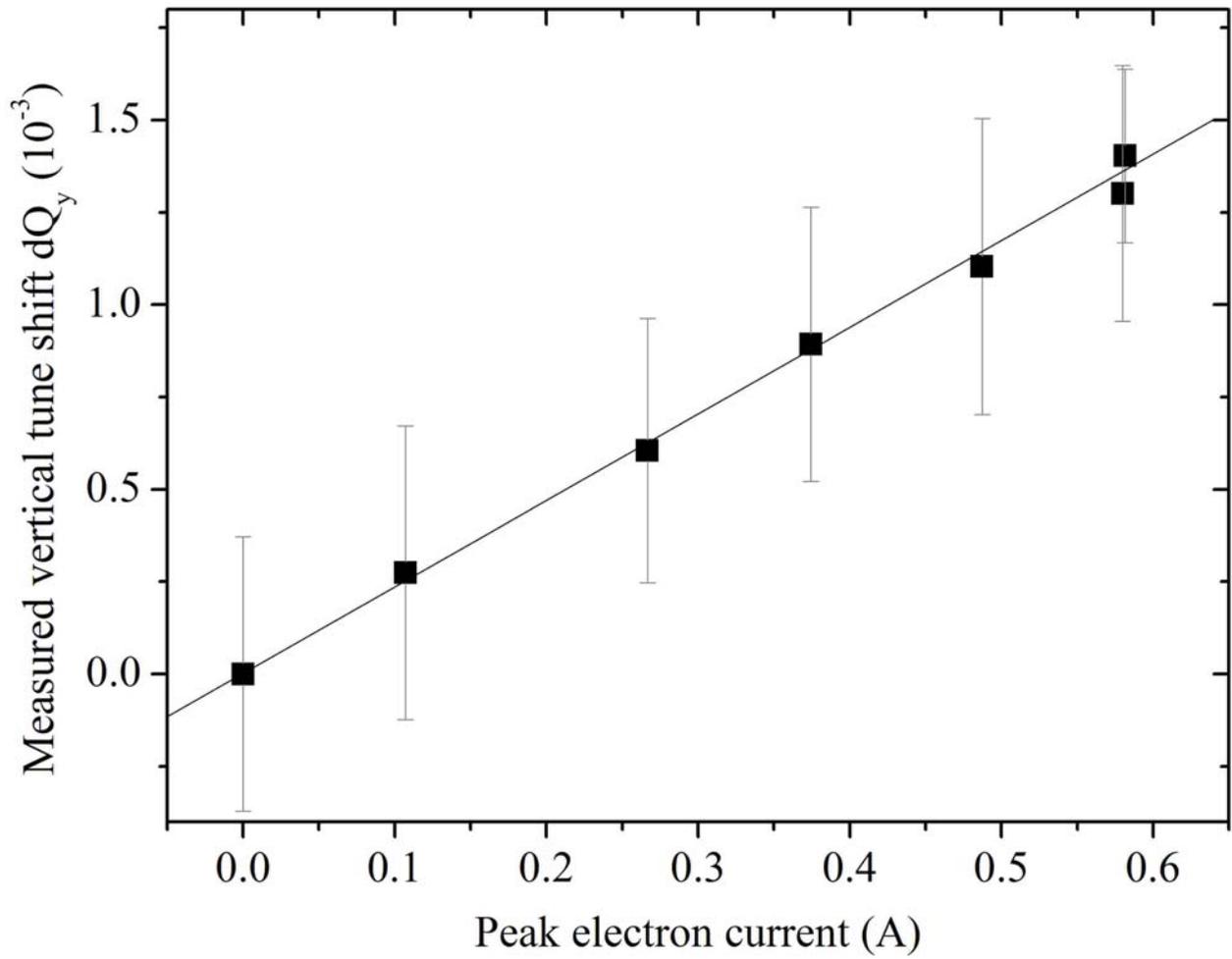

FIG.2. Vertical betatron tune shift of 980-GeV proton bunch vs. the peak electron current in the A11 TEL.

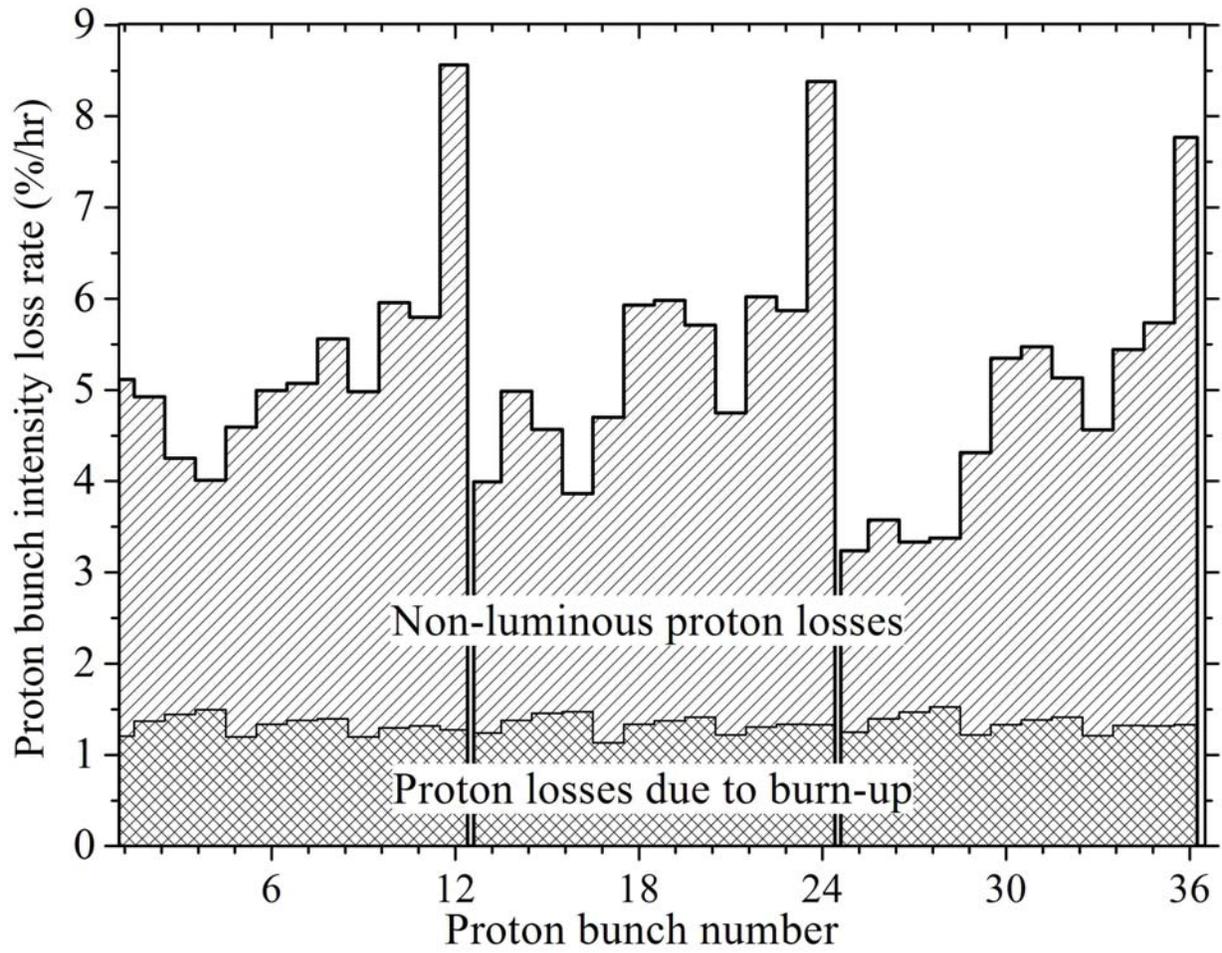

FIG.3. Proton-bunch intensity loss rates at the beginning of the Tevatron store #5155, Dec. 30, 2006, with initial luminosity $2.5 \cdot 10^{32}$ cm$^{-2}$ s$^{-1}$.

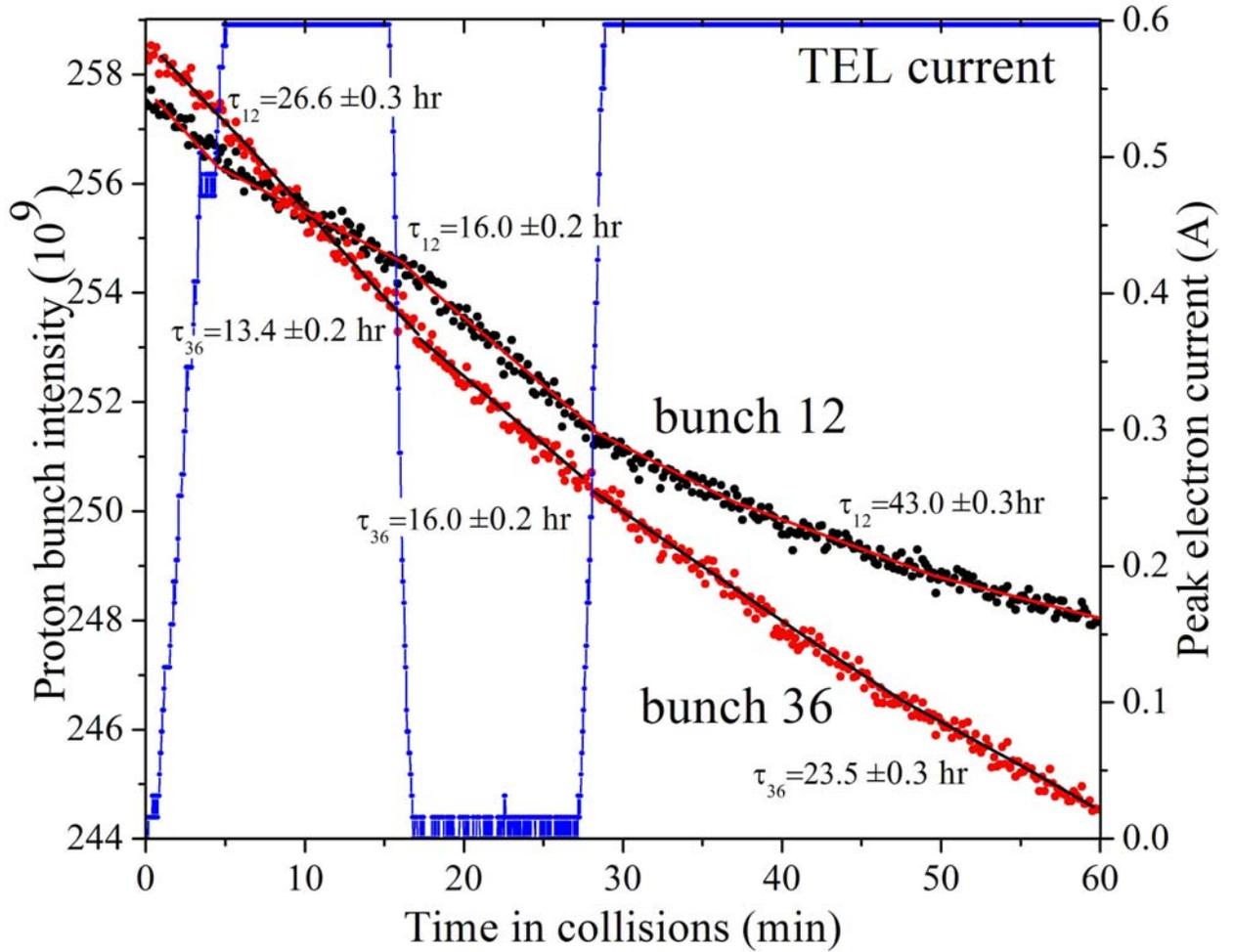

FIG.4. Intensities of proton bunches #12 and #36 during the first hour of the Tevatron store #5119 (Dec.12, 2006, initial luminosity $1.6 \cdot 10^{32}$ cm$^{-2}$ s$^{-1}$). The A11 TEL, timed only onto bunch #12, was turned on, off, and on.

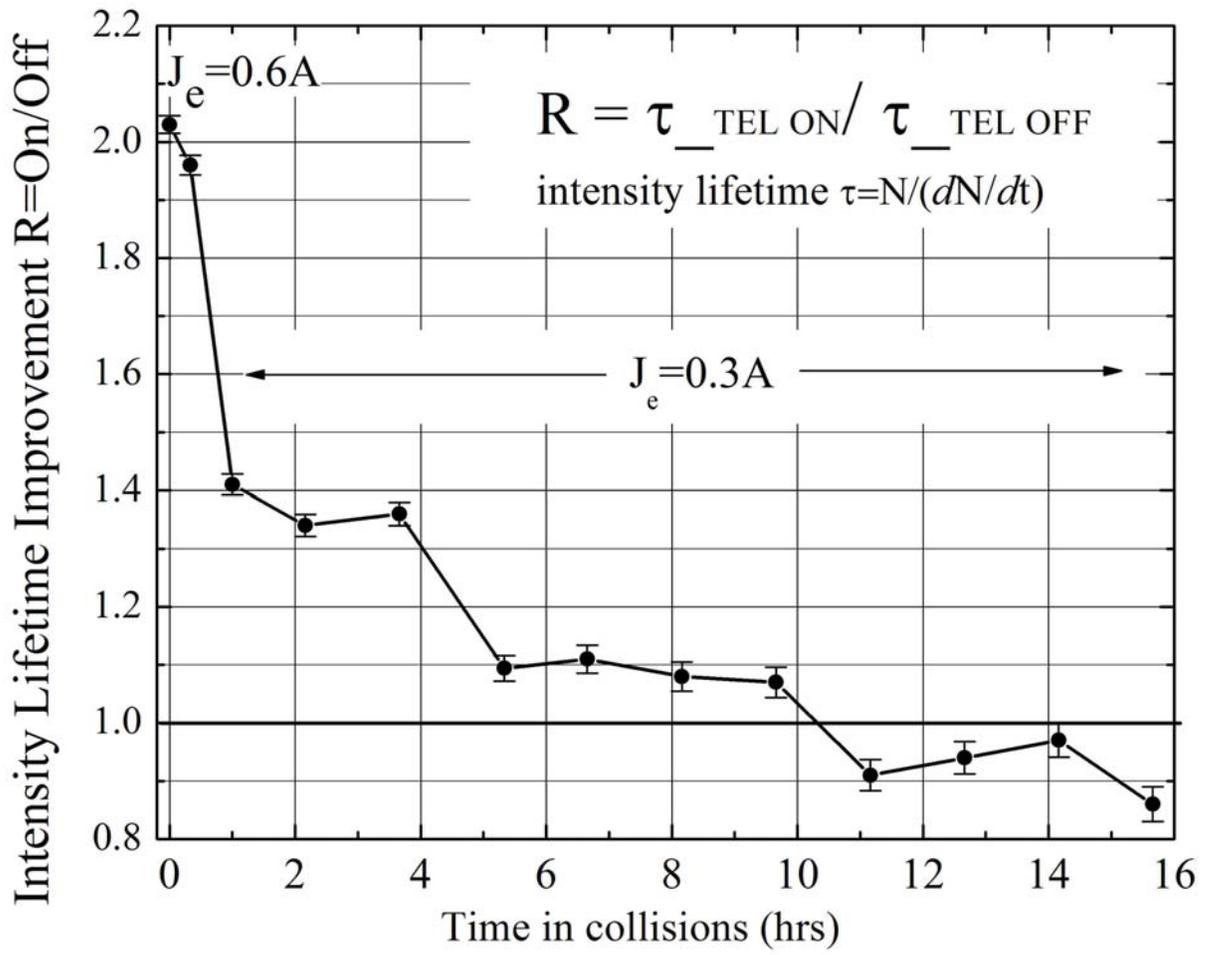

FIG.5. Relative improvement of the TEL induced proton bunch #12 lifetime vs. time (store #5119, Dec. 12, 2006).